\documentclass[referee]{raa}
\usepackage{graphicx,times}
\usepackage{natbib}
\usepackage{amssymb,amsmath}
\bibpunct{(}{)}{;}{a}{}{,}
\usepackage[a4paper=true,dvipdfm=true,pagebackref=true]{hyperref}
\hypersetup{pdftitle = The title of my PDF, pdfauthor = My name, pdfsubject= The subject, pdfkeywords = keyword1 keyword2 keyword3}
\hypersetup{colorlinks = true, linkcolor = green, anchorcolor = red, citecolor = blue, filecolor = red, pagecolor = red, urlcolor = red}
\begin{document}

   \title{Variable stars in the open cluster NGC 2141
$^*$
\footnotetext{\small $*$ Supported by the National Natural Science Foundation of China (11303021).}
}

 \volnopage{ {\bf 2014} Vol.\ {\bf X} No. {\bf XX}, 000--000}
   \setcounter{page}{1}

   \author{Yang-Ping Luo\inst{1,2}
   }

   \institute{ School of Physics \& Electronic Information, China West Normal University, Nanchong 637009,
China; {\it ypluo@bao.ac.cn}\\
        \and
             Key Laboratory for the Structure and Evolution of Celestial Objects, Chinese Academy of Sciences, Kunming 650011, China;\\
\vs \no
   {\small Received      ; accepted    }
}

\abstract{We report the results of a search for variable stars in the open cluster NGC 2141. Ten variable stars are detected, among which nine are
new variable stars and they are classified as three short-period W UMa-type eclipsing binaries, two EA-type eclipsing binaries, one EB-type eclipsing binary, one very short-period RS CVn-type eclipsing binary, one d-type RR Lyrae variable star, and one unknown type variable star. The membership and physical properties are discussed, based on their light curves, positions in the CMDs, spatial locations and periods. A known EB-type eclipsing binary is also identified as a blue struggler candidate of the cluster. Furthermore, we find that all eclipsing contact binaries have prominently asymmetric eclipses and O'Connell effect
 (\citealt{Connell1951}) which increases with the decrease of the orbital periods.
 This suggests that the O'Connell effect is probably related to the evolution of the orbital period in short period eclipsing binary systems.
}
   \authorrunning{Y.-P. Luo }            
   \titlerunning{Photometric study and search of variable stars in the open cluster NGC 2141}  
   \maketitle

\section{Introduction}           
\label{sect:intro}

This paper is a contribution to our ongoing program of a search for variable stars in open clusters(\citealt{Zhang2002, Zhang2004}; \citealt{luo2009, luo2012}).
One of our main goals is to study the stellar evolution via the census of variable stars in the open clusters.
The stars in an open cluster have a approximately same age and chemical abundances, in which the statistical properties of variable stars can put some stronger constraints on the stellar theoretical models (\citealt{Kim2000}) and the unresolved physical processes like the mass transfer, common-envelope ejection, mass loss, angular momentum loss, etc (\citealt{jiang2009}; \citealt{Pietrzynski2009, Pietrzynski2013}; \citealt{Ivanova2013}). In addition, they were also used to measure the age and distance of the hosted clusters and provide some hits on the cluster dynamical evolution. (\citealt{Meibom2009}).

NGC 2141 is an old galactic anticenter open cluster.
The coordinates of the cluster center are RA(J2000)=$06^{h}02^{m}58^{s}.2$ and
Dec(J2000)=$+10^{\circ}26'39''$ ($l=198^{\circ}.75$, $b=-5^{\circ}.79$) (\citealt{carraro2001}).
\cite{Burkhead1972} presented the first photoelectric and photographic observations and concluded that NGC 2141 is of a late intermediated age, a mean color excess of $E(B-V)=0.30$ and a distance modulus of $m-M=14.1$. \cite{rosvick1995} then performed a new optical $VI$ and near-infrared $JH$ photometry. Compared with the theoretical isochrones, the age and metallicity were estimated to be $t=2.5\,$Gyr and $Z=0.006$. The corresponding color excess and distance modulus were $E(B-V)=0.35\pm0.07$ and $m-M=14.16\pm0.16$.
\cite{carraro2001} acquired a deeper $BV$ and $JK$ photometry in a small area round the cluster center, from which they derived a slightly larger reddening ($E(B-V)=0.4$) and a slightly shorter distance ($\sim3.8$ kpc). The latest photometric data for this cluster have been presented by \cite{Donati2014}, who concluded that NGC 2141 is an old open cluster with age in the range 1.25 and 1.9 Gyr, $E(B-V)$ between 0.36 and 0.45, $(m-M)_{0}$ between 11.95 and 12.21 and subsolar metallicity. NGC 2141 is still an interesting target for the study of variable stars.
 Much deeper photometry showed that the cluster has a very rich binary population and appears the mass segregation (\citealt{carraro2001}; \citealt{Donati2014}). Although many variable candidates in NGC 2141 were reported, the detailed information is not yet known (\citealt{Kissinger2005}; \citealt{Widhalm2006}).
Therefore, we carried out a new time-series photometry for the open cluster NGC 2141 to detect the variable stars. Observations and data reductions are described in Section 2; identification of variable stars is presented in Section 3; their physical properties are discussed in Section 4; and a summery is followed in Section 5.

\section{Observations and Data Reductions}
\label{sect:Obs}
All photometric observations of the open cluster NGC 2141 were carried out on four nights in Jan 4--7, 2011,  using the Lijiang 2.4m telescope in Yunnan Astronomical Observatory, Chinese Academy of Sciences. We used the Yunnan Faint Object Spectrograph and Camera (YFOSC) to take the data, which are equipped with a 4096$\times$2048 CCD camera and two filter systems: standard johnson-cousin-bessel $UBVRI$ and SDSS $ugriz$. It provides two observational modes: spectra and image. We adopted the image mode, under which the effective pix is 2048$\times$2048 and the field of view is about $10'\times10'$. The Standard Johnson-Cousin-Bessel $V$ filter was chosen and the exposure times were set to be 360 second. In total, we acquired 230 $V$-band frames. An additional set of $BVRI$ photometry of the cluster was obtained in Jan 6, 2011, when it satisfied the photometric sky conditions. Two Landolt standard fields: SA 95 and SA 98 (\citealt{landolt1992}) and a Stetson photometric standard field: open cluster M67 (\citealt{stetson2000}) were also observed to construct photometric standards.

The raw images were de-biased and flat-fielded with the IRAF-CCDPROC package.
The instrumental magnitudes of stars in the CCD images were then extracted using
the point-spread function fitting program in the IRAF-DAOPHOT package.
The instrumental magnitudes in $BVRI$ bands were corrected by using IRAF-DAOGROW package and converted to standard system with the following transformation equations:

\begin{equation}
B=b-1.277\pm0.025+(0.031\pm0.004)(b-v)-(0.233\pm0.019)X,
 \end{equation}
\begin{equation}
V=v-1.052\pm0.013-(0.073\pm0.002)(b-v)-(0.119\pm0.010)X,
\end{equation}
\begin{equation}
R=r-0.991\pm0.009-(0.103\pm0.004)(v-r)-(0.095\pm0.007)X,
\end{equation}
\begin{equation}
I=i-0.624\pm0.024+(0.019\pm0.005)(v-i)-(0.065\pm0.018)X,
\end{equation}
where $b,v,r$ and $i$ are the instrumental magnitudes,
 $B,V,R$ and $I$  denote the standard magnitudes,
as well as $X$ is the airmass.

\section{Identification of Variable Stars}
\label{sect:iden}

For the purpose of searching for variable stars, we made a differential photometry for
each star detected in $V$-band images.
Followed \cite{Zhang2002}, an image with the best seeing and highest signal to noise ratio was chosen to be the reference frame, from which we then iteratively picked up about one hundred non-variable bright stars as the reference stars. With these reference stars, the magnitudes of all stars in images were corrected with respect to the reference frame.

\begin{figure}
\centering
\includegraphics[scale=0.7,bb=60 190 550 680]{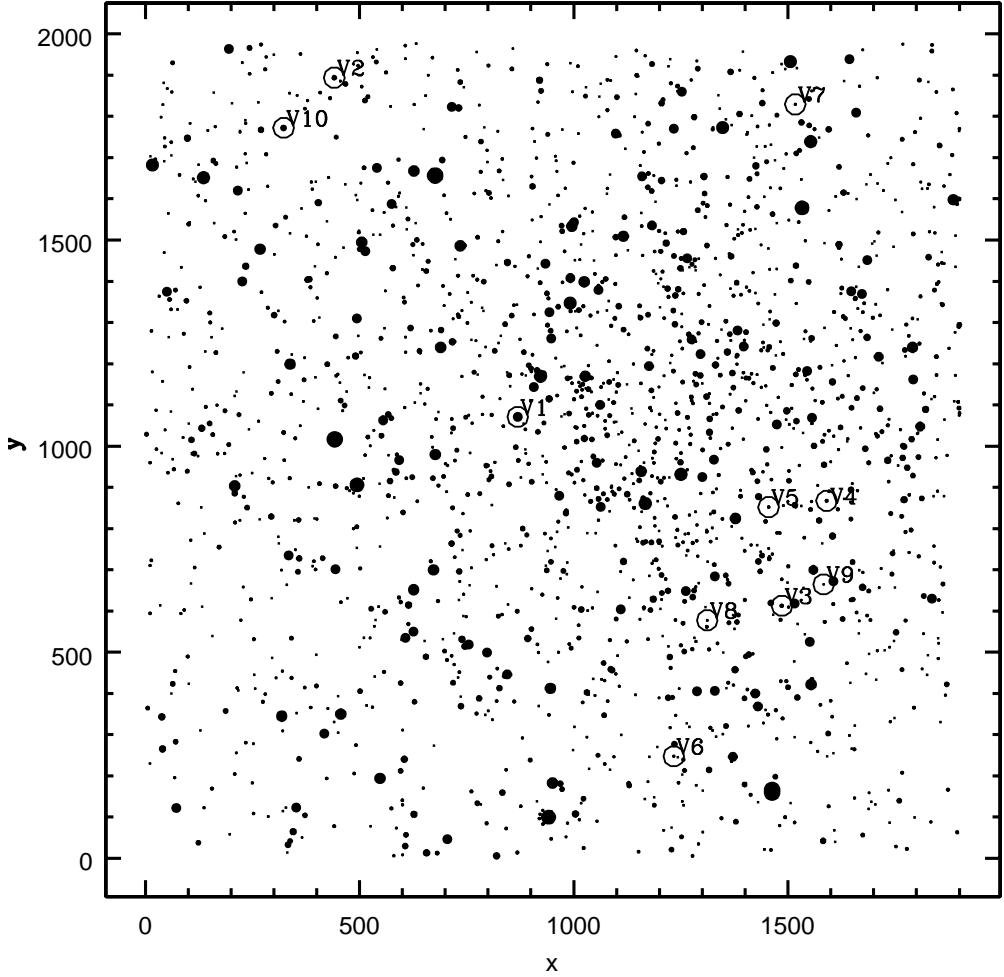}
\caption{Observed CCD field of the open cluster NGC 2141 and locations of variable stars}
\label{fig1}
\end{figure}

We then used two selection methods to identify variable candidates from above differential light curves.
Firstly, we selected the stars whose light curves show the larger deviations than those with similar brightness
as candidates. Then, we calculated the Stetson $J-$indexs (\citealt{stetson1996}) of stars and picked up stars with large $J-$index as candidates. Finally, we visually inspected the light curves of variable candidates and
rejected spurious variables and those showing small variability and
chaotic light variations. In total, we identified ten new variable stars in the field of NGC 2141. They are temporarily named as V1$-$V10. Figure.\ref{fig1} displays their spatial locations in the observed CCD field and
Table.\ref{tab1} gives their coordinates and physical parameters derived from the color magnitude diagrams (CMD) and light curves.

\begin{table}
\bc
\begin{minipage}[]{110mm}
\caption[]{Parameters of variable stars in the field of NGC 2141\label{tab1}}\end{minipage}
\setlength{\tabcolsep}{1pt}
\small
\begin{tabular}{clclclclclclclclc}
\hline\noalign{\smallskip}
Star & RA & Dec  & $V_{max}$ & $B-V$ & $V-R$  &  $V-I$ &     Radius &  Period & $T_{0}$ & Type & Memb  \\
  ID & (J2000)   & (J2000)    &   (mag)   &(mag)  &(mag)  & (mag) & (arcmin)  &(days)  & (days)  &      &       \\
\hline\noalign{\smallskip}
 V1& 06:03:03.85& 10:27:15.68& 15.046& 0.604& 0.408& 0.924&  1.78 & 0.6233& 68.1022& EB & likely\\
 V2& 06:03:12:33& 10:31:09:82& 16.857& 0.820& 0.528& 1.049&  5.42 & 0.3984& 70.1991& RRd& unlikely \\
 V3& 06:02:51.90& 10:25:07.05& 17.230& 0.894& 0.561& 1.084&  2.52 & 1.3665& 68.2472& EA & likely\\
 V4& 06:02:49.37& 10:26:14.71& 17.981& 0.872& 0.536& 1.054&  1.95 & 4.9878& 70.3432& EA & likely\\
 V5& 06:02:52.54& 10:26:15.50& 18.213& 0.987& 0.589& 1.157&  1.49 & 0.3261& 67.3106& EW & likely\\
 V6& 06:02:55.28& 10:24:58.06& 18.264& 0.924& 0.579& 1.193&  3.98 & 0.3112& 67.1370& EW & likely\\
 V7& 06:02:51.52& 10:30:54.07& 19.031& 0.975& 0.663& 1.275&  3.79 & 0.5529& 70.1531& EB & likely\\
 V8& 06:02:56.72& 10:23:22.77& 19.111& 1.134& 0.723& 1.466&  2.44 & 0.2305& 70.1278& RS CVn & likely\\
 V9& 06:02:50.20& 10:25:22.10& 19.231& 1.362& 0.943& 1.712&  2.59 & 0.2432& 70.0987& EW & unlikely\\
V10& 06:03:14.57& 10:30:34.05& 16.120& 1.056& 0.663& 1.329&  5.48 &      &        &    &likely\\
\hline
\end{tabular}
\ec
\tablecomments{0.86\textwidth}{$T_{0}(HJD-2455500)$ denotes the phase zero epoch.}
\end{table}

\section{Physical Properties of Variable stars}
\label{sect:prop}
\subsection{Cluster membership}

In general, the determination of the cluster membership mainly depends on the proper motions and
radial velocities. However, there is not any report on them. In this paper, we discuss their cluster memberships only according to their spatial locations in the cluster field and positions in the CMDs.

\begin{figure}
\centering
\includegraphics[scale=0.6,bb=60 180 400 500]{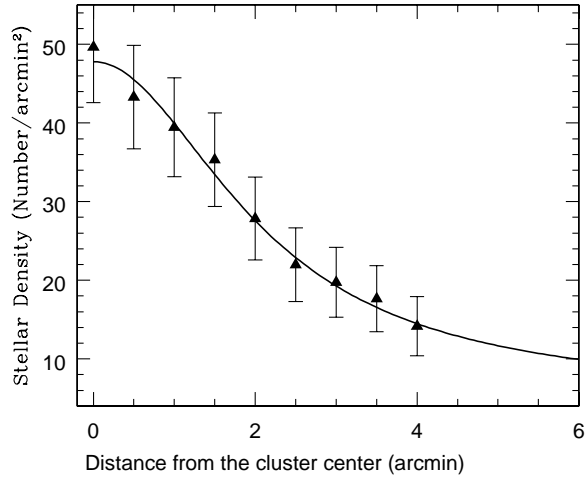}
\caption{The stellar radial density profile for stars brighter than 20.0 mag in the field of NGC 2141. The sold line denotes the King profile.}
\label{fig2}
\end{figure}

To derive the cluster memberships of variable stars from the spatial locations, we derived the physical parameters of the cluster. First of all, we determined the center of the clusters
by finding the maximum surface number density of stars in the cluster filed. Here, stars with $V\leq20$ were just considered.
The cluster center was roughly determined by the pixel coordinate (1234,1080) in our reference CCD frame. Then, we set a series of concentric rings around the center. The width of rings was set to be 106 pix($\sim30$ arcsec). The stellar radial density profile was derived by counting the number of stars per area in each concentric ring and is shown in Fig.\ref{fig2}. The error bars were determined on the assumption that the number of stars in each rings follows the Poisson statistical distribution. We adopted a two-parameter King model (\citealt{King1966}) to fit the radial density profile
\begin{equation}
\sigma(r)=\sigma_{bg}+\frac{\sigma_{0}}{1+(\frac{r}{r_{c}})^2},
\end{equation}
where $\sigma_{bg}$ is the background field
density, $\sigma_{o}$ the central density of stars, and $r_{c}$ the core
radius of the cluster.  A best fitting model is shown by the sold line in Fig.\ref{fig2} and gives that $\sigma_{bg}=5.3\pm3.2$ stars arcmin$^{-2}$, $\sigma_{0}=42.5\pm3.5$ stars arcmin$^{-2}$ and $r_{c}=2.1\pm0.3$ arcmin.
The distances of variable stars from the center are given in Table.\ref{tab1}.
We could deduce from Fig.\ref{fig2} that our observed filed is inside the cluster, which implies that all variable stars are probably the cluster members in the spatial locations.

\begin{figure}
\centering
\includegraphics[width=0.9\textwidth,bb=60 280 582 590]{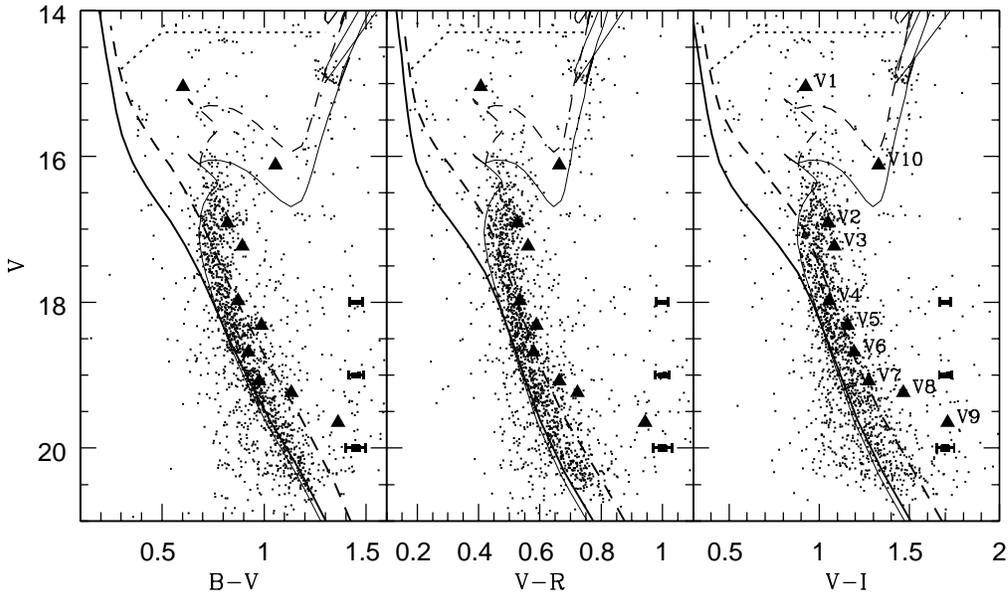}\vspace{-2mm}
\caption{CMDs of NGC 2141 and positions of variable stars in the CMDs. The thin solid lines denote the fit of the Padova theoretical isochrones with an age of $t=1.9\,$Gyr, a metallicity of $Z=0.008$, an reddening of $E(B-V)=0.36\,$mag, and  a distance modulus of $(m-M)_{0}=13.20\,$mag (\citealt{Donati2014}) and the thick solid lines are the zero age main sequences. The dashed lines are the corresponding binary sequences. The dotted lines approximately indicate the upper edge of blue stragglers (\citealt{lu2010})}.
\label{fig3}
\end{figure}

In addition, CMDs can also provide some very important constraints on the cluster membership of the variable stars.
CMDs of the NGC 2141 were constructed with $BVRI$ multi-color photometric data and are displayed in Fig.\ref{fig3}. We mark their positions on the CMDs and plot the Padova theoretical isochrones (\citealt{Girardi2002}) with the cluster physical parameters (age: $t=1.9\,$Gyr, metallicity: $Z=0.008$, reddening: $E(B-V)=0.36\,$mag, distance modulus: $(m-M)_{0}=13.20\,$mag) derived by \cite{Donati2014}.
Generally, stars in the open cluster distribute along with the isochrones.  Two variable stars (V1 and V9) are far away from the isochrones. However, it is noted that the position of V1 in the CMDs settles in the main sequence but brighter than the turn-off stars. The current observations and binary evolution theories (\citealt{bailyn1995};\citealt{lu2010};\citealt{geller2011}) showed that the cluster members are allow to locate at the area of V1. As a result, we concluded that all variable stars except for V9 are probably the cluster members.

\subsection{Periodicity analysis}

\begin{figure}
\centering
\includegraphics[width=0.9\textwidth,bb=50 300 552 690]{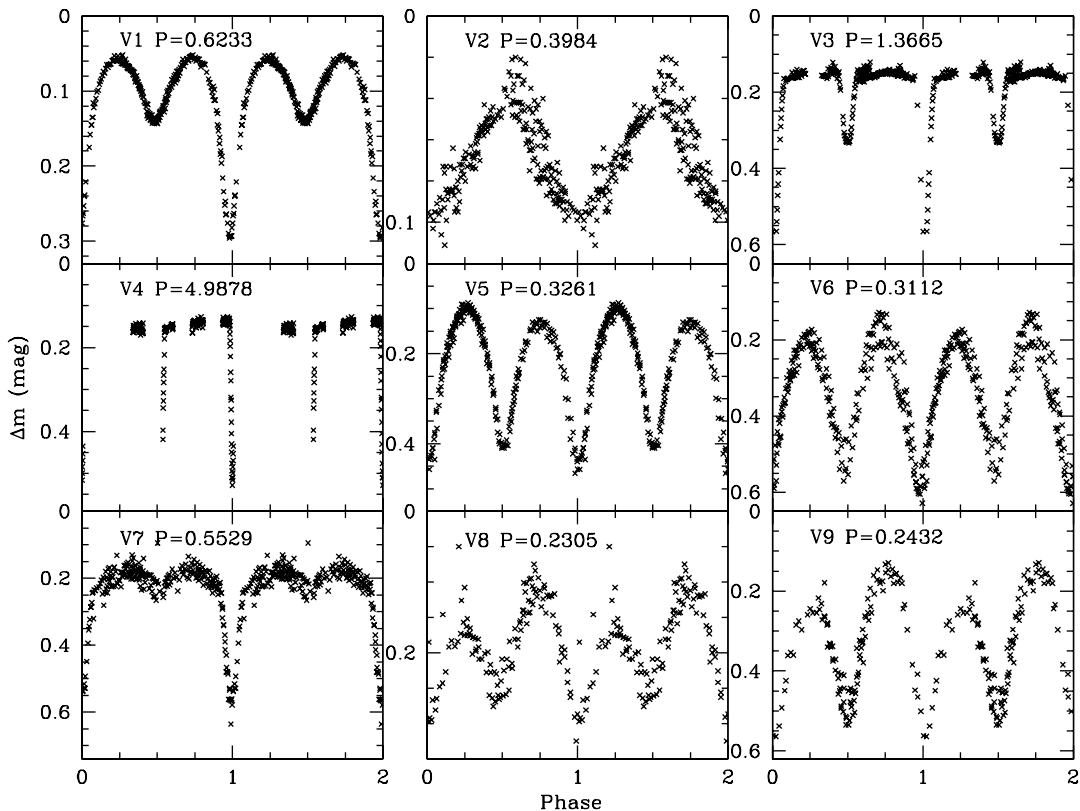}\vspace{-2mm}
\caption{Phased light curves of nine periodic variable star in the field of NGC 2141.} 
\label{fig4}
\end{figure}

\begin{figure}
\centering
\includegraphics[scale=0.45]{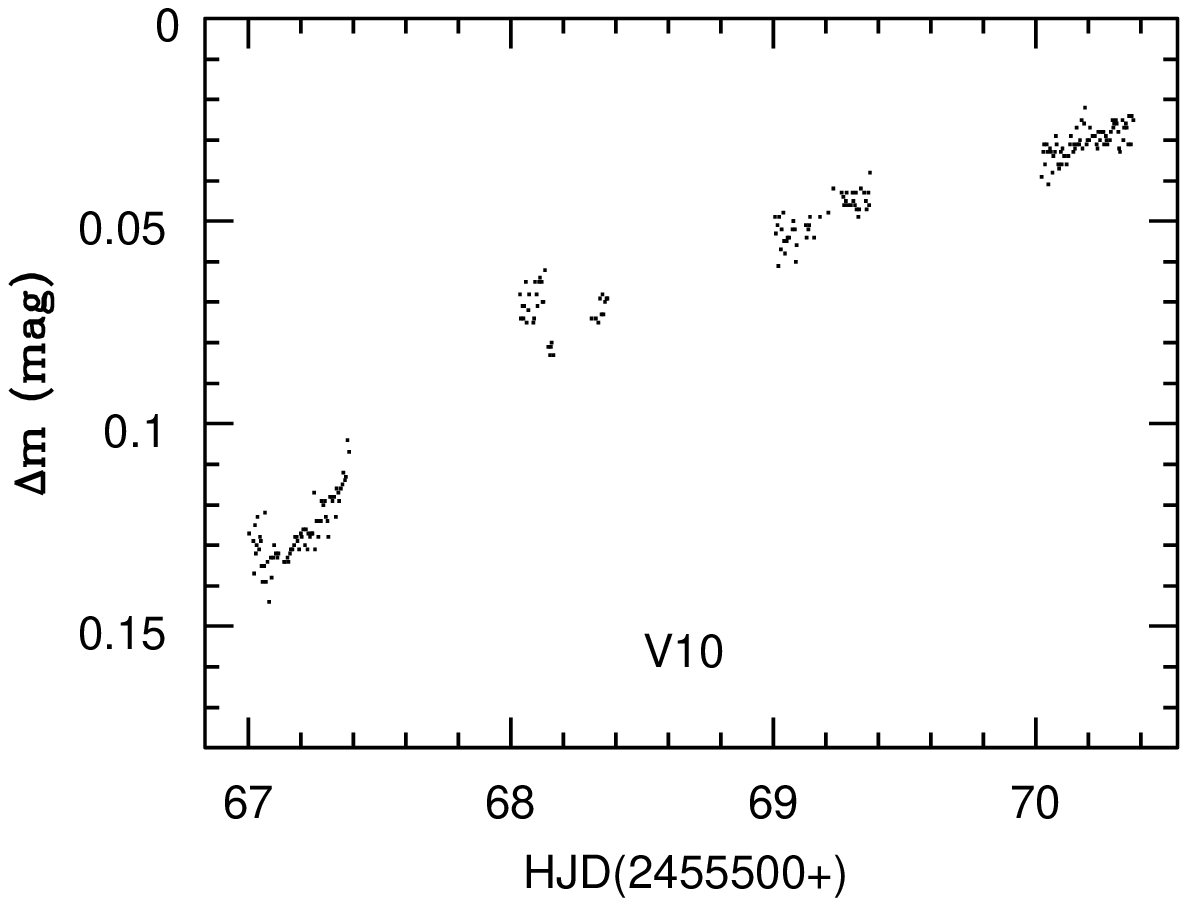}\vspace{-2mm}
\caption{Light curves of V10, an unknown type variable star in field of NGC 2141.}
\label{fig5}
\end{figure}

We used the PDM program in the IRAF-ASTUTIL package to determine the periods of variable stars, which
is based on the phase dispersion minimization algorithm method(\citealt{stellingwerf1978}).
To do this, we set a range of period to be 0.01-8 days. Then, we used the PDM program to estimate a few possible periods in the range and derive the corresponding phase-folded light curves. After visually inspecting the phase-folded light curves, we determined a best ones. Nine stars(V1-V9) were found to be the periodic variable stars and their periods and phase zero points are given in Table.\ref{tab1}. The phase-folded light curves are shown in Fig.\ref{fig4}.
The light curves of V10 is shown in Fig.\ref{fig5}, in which the periodical variability is not found in our observations.

\subsection{Classification and Discussions}

We classified the variable stars, mainly based on the
shapes of light curves, the detected periods and the positions in the CMDs.
The classification and characterization of the variable stars in our study are discussed as follows.

  V1 has been classified as an EB-type eclipsing binary star by \cite{Licchelli2011} and was named as VSX J060303.8+102715 in American Association of Variable Star Observers (AAVSO)\footnote{http://www.aavso.org.}. It is a likely cluster member and the period was updated as 0.6233days. It is very interesting that V1 is more probably a blue straggler candidate of the cluster. The position of V1 in the CMDs is just in the region where the formation of the blue stragglers via the mass transfer have predicted (\citealt{chen2008}; \citealt{lu2010}) and observationally many blue stragglers were found (\citealt{geller2011}). Our detailed photometric solution (\citealt{luo2012b}) showed that V1 is a semi-detached binary with a less massive component that is filling the Roche Lobe and is super-luminance.
 Here, the discovery of V1 is prefer to support a conclusion that the blue stragglers in the open cluster origin from the mass transfer of binary stars (\citealt{geller2011}).
Moreover, V1 is a rare eclipsing blue straggler binary system of the cluster that is transferring the mass form the less massive component to the massive ones (\citealt{luo2012b}). Therefore, follow-up observations are important for investigating the physical process of the mass transfer during the formation of blue stragglers of the open cluster.
The light curve of V7 is similar to V1. It is identified as an EB-type eclipsing binary star and the period was determined as 0.5529 days. In the CMDs, this star is located at the main sequence of the cluster binary stars. This star is also a likely cluster member.

V2 shows a peculiar light curve with slowly descending and quick ascending branches.
The shape of light curves is similar to the RR Lyrae variable star.
The most probable period was determined to be 0.3984 days, which is too long for
normal $\delta$ Scuti star. Therefore, it is reasonable that V2 is classified as an RR Lyrae variable star.
From Fig.$\,$4, we can clearly see that
the light curve appears obviously high asymmetry and rapid changes from one period to another.
The period modulations can also be seen in phased light curves. These characteristics suggest that V2 is a d-type RR Lyrae star. The position in the CMDs is just on the main sequence, therefore, we could conclude that V2 is not the cluster member but a background filed star.

V3 and V4 have a flat light maxima and are identified as the EA-type eclipsing binaries.
They are the probable cluster member. The periods of V3 is estimated as 1.3665 days. However,
the light curve of V4 was incomplete and the period is roughly determined to be 4.9878 days. Further observations are needed to ascertain the exact nature.

There are four short period eclipsing contact binary systems (V5, V6, V8 and V9) with the orbital period of $0.23 \sim 0.33$ days. The characteristics of their light curves are very similar. The light curves have a clear difference in depths of eclipses, in which the primary eclipses are deeper than secondary ones. Moreover, the light curves show the O'Connell effect (\citealt{Connell1951}). The secondary light maxima are lager than the primary ones for V6, V8 and V9, on the contrary for V5. The characteristic parameters of the light curves are given in Table.\ref{tab2}, from which we could find that the O'Connell effect increases with the decrease of the orbital period. This shows that the O'Connell effect is probably related to the evolution of the orbital period in short period eclipsing binary systems.
The O'Connell effect of V8 is stronger than others, even more than 63 percent of the amplitude of light curve. Therefore,  it is identified as the RS CVn-type eclipsing binary system. The others( V5, V6 and V9) have the general nature of the W Uma-type eclipsing binary system and are classified as the W Uma binary.
However, the O'Connell effect in short period eclipsing binary systems is still uncertain. Many different theories have been proposed to explain to this effect, but no one theory have been successfully applied to more than a handful of binary systems (\citealt{Wilsey2009}). Further multi-color photometry and spectroscopy observations may put some insights into the nature of the O'Connell effect in eclipsing binary system.

V10 is classified as an unknown type variable star. The time-series light curves does not show any periodical variability in our observations. Further observations will help us determine their periods and physical properties.

\begin{table}
\begin{center}
\caption[]{Characteristic parameters of the light curves of four eclipsing contact binaries (V5, V6, V8, V9)
$\Delta (MinI-MinII)$ is the difference of primary and secondly eclipse, $\Delta (MaxI-MaxII)$ is the difference of primary and secondly light maxima and  $A$ is the amplitude of light curve.

}\label{tab2}
 \begin{tabular}{clclclclclcl}
  \hline\noalign{\smallskip}
Star &  $\Delta (MinI-MinII)$ & $\Delta (MaxI-MaxII)$ & $A \times 2$      & $\Delta (MaxI-MaxII)/A$  & period \\
     &   (mag)            &   (mag)           &  (mag)   &  ($\%$)              & days     \\
\hline\noalign{\smallskip}
V5   &   0.06   & 0.04  &  0.38  &  21.1 & 0.3261  \\
V6   &   0.06   & 0.05  &  0.47  &  21.3 & 0.3112  \\
V8   &   0.03   & 0.07  &  0.22  &  63.6 & 0.2305  \\
V9   &   0.02   & 0.09  &  0.23  &  41.9 & 0.2432 \\
  \noalign{\smallskip}\hline
\end{tabular}
\end{center}
\end{table}

\section{summery}
\label{sect:discussion}

In this paper, we have presented a time-series $V$band and muti-color $BVRI$ CCD photometry for the open
cluster NGC 2141, undertaken in 2011 to detect variable stars.
The following conclusions can be drawn:
\begin{enumerate}

\item Ten variable stars have been detected in the
field of the old open cluster NGC 2141, among which
nine are newly discovered.
We discussed their memberships on the basis of their space locations, positions in the CMDs, and physical properties.
Seven stars (V1, V3, V4, V5, V6, V7, V8 and V10) are the probable members of the cluster, while
the others (V2 and V9) are the unlikely cluster members. We found that nine stars are the periodic variable stars and their periods were determined with the phase dispersion minimization
algorithm method.

\item We assessed the classifications of variable stars and discussed the physical properties, primarily based on the shape of light curve, the detected period and the position in the CMDs.
They are categorized as three W UMa-type eclipsing binaries, two EA-type eclipsing binaries,
 two EB-type eclipsing binaries, one d-type RR Lyre star, one RS CVn-type eclipsing binary and one unknown type variable star. A known EB-type eclipsing binary V1 settles in the area of the cluster blue straggler in the CMDs and it is identified to be a blue straggler candidate.

 \item In this study, we found that all four short period eclipsing contact binaries show a clear difference in depths of eclipses and O'Connell effect which increases with the decrease of the orbital periods. This illustrates that the O'Connell effect is probably related to the evolution of the orbital period in short period eclipsing binary systems.

\end{enumerate}

\normalem
\begin{acknowledgements}
We thank the referee for a detailed review and helpful suggestions.
This work was supported by the National
Natural Science Foundation of China  (11303021)
and the Lijiang 2.4m telescope, Yunnan Observatories, CAS.
\end{acknowledgements}



\begin{thebibliography}{27}
\providecommand{\natexlab}[1]{#1}
\providecommand{\selectlanguage}[1]{\relax}

\bibitem[{{Bailyn}(1995)}]{bailyn1995}
{Bailyn}, C.~D. 1995, \araa, 33, 133

\bibitem[{{Burkhead} et~al.(1972){Burkhead}, {Burgess}, \&
  {Haisch}}]{Burkhead1972}
{Burkhead}, M.~S., {Burgess}, R.~D., \& {Haisch}, B.~M. 1972, \aj, 77, 661

\bibitem[{{Carraro} et~al.(2001){Carraro}, {Hassan}, {Ortolani}, \&
  {Vallenari}}]{carraro2001}
{Carraro}, G., {Hassan}, S.~M., {Ortolani}, S., \& {Vallenari}, A. 2001, \aap,
  372, 879

\bibitem[{{Chen} \& {Han}(2008)}]{chen2008}
{Chen}, X., \& {Han}, Z. 2008, \mnras, 384, 1263

\bibitem[{{Donati} et~al.(2014){Donati}, {Beccari}, {Bragaglia}, {Cignoni}, \&
  {Tosi}}]{Donati2014}
{Donati}, P., {Beccari}, G., {Bragaglia}, A., {Cignoni}, M., \& {Tosi}, M.
  2014, \mnras, 437, 1241

\bibitem[{{Geller} \& {Mathieu}(2011)}]{geller2011}
{Geller}, A.~M., \& {Mathieu}, R.~D. 2011, \nat, 478, 356

\bibitem[{{Girardi} et~al.(2002){Girardi}, {Bertelli}, {Bressan}
  et~al.}]{Girardi2002}
{Girardi}, L., {Bertelli}, G., {Bressan}, A., et~al. 2002, \aap, 391, 195

\bibitem[{Kissinger} \& {Kafka}(2005)]{Kissinger2005}
{Kissinger}, J., \& {Kafka}, S. 2005, American Astronomical Society Meeting Abstracts, 37, 1278


\bibitem[{{Ivanova} et~al.(2013){Ivanova}, {Justham}, {Chen}
  et~al.}]{Ivanova2013}
{Ivanova}, N., {Justham}, S., {Chen}, X., et~al. 2013, \aapr, 21, 59

\bibitem[{{Jiang} et~al.(2009){Jiang}, {Han}, {Jiang}, \& {Li}}]{jiang2009}
{Jiang}, D., {Han}, Z., {Jiang}, T., \& {Li}, L. 2009, \mnras, 396, 2176

\bibitem[{{Kim} et~al.(2000){Kim}, {Park}, {Chun} et~al.}]{Kim2000}
{Kim}, S.-L., {Park}, B.-G., {Chun}, M.-Y., et~al. 2000, 203, 457

\bibitem[{{King}(1966)}]{King1966}
{King}, I.~R. 1966, \aj, 71, 64

\bibitem[{{Landolt}(1992)}]{landolt1992}
{Landolt}, A.~U. 1992, \aj, 104, 340

\bibitem[{{Licchelli}(2011)}]{Licchelli2011}
{Licchelli}, D. 2011, http://www.aavso.org/vsx/index.php?view=detail.top=256292

\bibitem[{{Lu} et~al.(2010){Lu}, {Deng}, \& {Zhang}}]{lu2010}
{Lu}, P., {Deng}, L.~C., \& {Zhang}, X.~B. 2010, \mnras, 409, 1013

\bibitem[{{Luo} et~al.(2012{\natexlab{a}}){Luo}, {Wang}, {Chen}, {Zhang}, \&
  {Han}}]{luo2012b}
{Luo}, Y., {Wang}, B., {Chen}, H., {Zhang}, X., \& {Han}, Z.
  2012{\natexlab{a}}, Science China Physics, Mechanics, and Astronomy, 55, 1500

\bibitem[{{Luo} et~al.(2012{\natexlab{b}}){Luo}, {Zhang}, {Deng}, \&
  {Han}}]{luo2012}
{Luo}, Y.~P., {Zhang}, X.~B., {Deng}, L.~C., \& {Han}, Z.~W.
  2012{\natexlab{b}}, \apjl, 746, L7

\bibitem[{{Luo} et~al.(2009){Luo}, {Zhang}, {Luo}, {Deng}, \& {Luo}}]{luo2009}
{Luo}, Y.~P., {Zhang}, X.~B., {Luo}, C.~Q., {Deng}, L.~C., \& {Luo}, Z.~Q.
  2009, \na, 14, 584

\bibitem[{{Meibom} et~al.(2009){Meibom}, {Grundahl}, {Clausen}
  et~al.}]{Meibom2009}
{Meibom}, S., {Grundahl}, F., {Clausen}, J.~V., et~al. 2009, \aj, 137, 5086

\bibitem[{{O'Connell}(1951)}]{Connell1951}
{O'Connell}, D.~J.~K. 1951, Publications of the Riverview College Observatory,
  2, 85

\bibitem[{{Pietrzy{\'n}ski} et~al.(2013){Pietrzy{\'n}ski}, {Graczyk}, {Gieren}
  et~al.}]{Pietrzynski2013}
{Pietrzy{\'n}ski}, G., {Graczyk}, D., {Gieren}, W., et~al. 2013, \nat, 495, 76

\bibitem[{{Pietrzy{\'n}ski} et~al.(2009){Pietrzy{\'n}ski}, {Thompson},
  {Graczyk} et~al.}]{Pietrzynski2009}
{Pietrzy{\'n}ski}, G., {Thompson}, I.~B., {Graczyk}, D., et~al. 2009, \apj,
  697, 862

\bibitem[{{Rosvick}(1995)}]{rosvick1995}
{Rosvick}, J.~M. 1995, \mnras, 277, 1379

\bibitem[{{Stellingwerf}(1978)}]{stellingwerf1978}
{Stellingwerf}, R.~F. 1978, \apj, 224, 953

\bibitem[{{Stetson}(1996)}]{stetson1996}
{Stetson}, P.~B. 1996, \pasp, 108, 851

\bibitem[{{Stetson}(2000)}]{stetson2000}
{Stetson}, P.~B. 2000, \pasp, 112, 925

\bibitem[{Widhalm} \& {Kafka}(2006)]{Widhalm2006}
{Widhalm}, A.~M., \& {Kafka}, S. 2006, American Astronomical Society Meeting Abstracts, 38, 1128



\bibitem[{{Wilsey} \& {Beaky}(2009)}]{Wilsey2009}
{Wilsey}, N.~J., \& {Beaky}, M.~M. 2009, Society for Astronomical Sciences
  Annual Symposium, 28, 107

\bibitem[{{Zhang} et~al.(2002){Zhang}, {Deng}, {Tian}, \& {Zhou}}]{Zhang2002}
{Zhang}, X.~B., {Deng}, L., {Tian}, B., \& {Zhou}, X. 2002, \aj, 123, 1548

\bibitem[{{Zhang} et~al.(2004){Zhang}, {Deng}, {Zhou}, \& {Xin}}]{Zhang2004}
{Zhang}, X.~B., {Deng}, L., {Zhou}, X., \& {Xin}, Y. 2004, \mnras, 355, 1369

\end{thebibliography}
\end{document}